\documentclass[twocolumn,
superscriptaddress,
showpacs,preprintnumbers,
 amsmath,amssymb,
 aps,
prb,
]{revtex4-1}

\usepackage{graphicx}
\usepackage{dcolumn}
\usepackage{bm}

\begin{document}

\preprint{APS}

\title{Local spin structure of the $\alpha$-RuCl$_3$ honeycomb-lattice magnet \\ observed via muon spin rotation/relaxation}

\author{Ichihiro Yamauchi}\email{ichihiro@cc.saga-u.ac.jp}
\affiliation{Department of Physics, Graduate School of Science and Engineering, Saga University, Saga 840-8502, Japan}
\author{Masatoshi Hiraishi}
\affiliation{Muon Science Laboratory and Condensed Matter Research Center, Institute of Materials Structure Science, High Energy Accelerator Research Organization (KEK), Tsukuba, Ibaraki 305-0801, Japan}
\author{Hirotaka Okabe}
\affiliation{Muon Science Laboratory and Condensed Matter Research Center, Institute of Materials Structure Science, High Energy Accelerator Research Organization (KEK), Tsukuba, Ibaraki 305-0801, Japan}
\author{Soshi Takeshita}
\affiliation{Muon Science Laboratory and Condensed Matter Research Center, Institute of Materials Structure Science, High Energy Accelerator Research Organization (KEK), Tsukuba, Ibaraki 305-0801, Japan}
\author{Akihiro Koda}
\affiliation{Muon Science Laboratory and Condensed Matter Research Center, Institute of Materials Structure Science, High Energy Accelerator Research Organization (KEK), Tsukuba, Ibaraki 305-0801, Japan}
\affiliation{Department of Materials Structure Science, The Graduate University for Advanced Studies (Sokendai), Tsukuba, Ibaraki 305-0801, Japan}
\author{Kenji M. Kojima}
\affiliation{Muon Science Laboratory and Condensed Matter Research Center, Institute of Materials Structure Science, High Energy Accelerator Research Organization (KEK), Tsukuba, Ibaraki 305-0801, Japan}
\affiliation{Department of Materials Structure Science, The Graduate University for Advanced Studies (Sokendai), Tsukuba, Ibaraki 305-0801, Japan}
\author{Ryosuke Kadono}
\affiliation{Muon Science Laboratory and Condensed Matter Research Center, Institute of Materials Structure Science, High Energy Accelerator Research Organization (KEK), Tsukuba, Ibaraki 305-0801, Japan}
\affiliation{Department of Materials Structure Science, The Graduate University for Advanced Studies (Sokendai), Tsukuba, Ibaraki 305-0801, Japan}
\author{Hidekazu Tanaka}
\affiliation{Department of Physics, Tokyo Institute of Technology, Meguro-ku, Tokyo 152-8551, Japan}

\date{\today}

\begin{abstract}

We report a muon spin rotation/relaxation ($\mu$SR) study of single-crystalline samples of the $\alpha$-RuCl$_3$ honeycomb magnet, which is presumed to be a model compound for the Kitaev-Heisenberg interaction. It is inferred from magnetic susceptibility and specific-heat measurements that the present samples exhibit successive magnetic transitions at different critical temperatures $T_{\rm N}$ with decreasing temperature, eventually falling into the $T_{\rm N}=7$ K antiferromagnetic (7 K) phase that has been observed in only single-crystalline specimens with the least stacking fault. Via $\mu$SR measurements conducted under a zero external field, we show that such behavior originates from a phase separation induced by the honeycomb plane stacking fault, yielding multiple domains with different $T_{\rm N}$'s. We also perform $\mu$SR measurements under a transverse field in the paramagnetic phase to identify the muon site from the muon-Ru hyperfine parameters.  Based on a comparison of the experimental and calculated internal fields at the muon site for the two possible spin structures inferred from neutron diffraction data, we suggest a modulated zig-zag spin structure for the 7 K phase, with the amplitude of the ordered magnetic moment being significantly reduced from that expected for the orbital quenched spin-1/2 state.  

\end{abstract}

\maketitle


\section{Introduction}

 The Kitaev model, a spin-1/2 honeycomb model with bond-dependent exchange interaction, has attracted considerable attention, because of its remarkable prediction of the spin-liquid state with fractional fermionic excitations. \cite{Kitaev2006} It is predicted that such a bond-dependent exchange interaction $K$ can be realized in a honeycomb magnet with effective spin $J_{\rm eff} = 1/2$ because of the strong spin-orbit coupling (SOC). \cite{Jackeli2009} The Kitaev-Heisenberg (KH) model, which is regarded as a more realistic model, comprises two additional interactions, i.e., the isotropic exchange interaction $J$ and the bond-dependent symmetric off-diagonal exchange interaction $\Gamma$. The KH model predicts an extended spin-liquid phase and several non-trivial antiferromagnetic (AFM) states upon tuning of the $K$, $J$, and $\Gamma$ magnitudes. \cite{Chaloupka2010,Rau2014} As a model compound for the KH model, $\alpha$-Na$_2$IrO$_3$ has been successfully synthesized, and a zig-zag type magnetic ordering has been found to be the ground state. \cite{Singh2010,Liu2011,Ye2012,Choi2012} However, the localized $J_{\rm eff} = 1/2$ picture remains under debate. \cite{Mazin2012,Foyevtsova2013}   

$\alpha$-RuCl$_3$ has attracted renewed interest as another candidate compound for the KH model. This compound crystallizes in a monoclinic structure with a $C2/m$ space group. \cite{Johnson2015,Cao2016} The RuCl$_6$ octahedra form an almost ideal honeycomb structure in the $ab$ plane by sharing the edges. Strong in-plane magnetic anisotropy observed via magnetization measurements suggests that the Ru$^{3+}$ $(4d^5)$ ion has a low-spin state. \cite{Kubota2015,Sears2015,Majumder2015,Johnson2015} Consequently, the ground state may be described as having an effective angular momentum $l=1$ with a total spin $S=1/2$; thus, the magnetic moment is given by $J_{\rm eff} = 1/2$. Indeed, Raman and neutron scattering experiments have revealed an excitation between the $J_{\rm eff} = 1/2$ ground state and $J_{\rm eff} = 3/2$ excited state. \cite{Sandilands2016,Banerjee2016} Various experimental studies support gap opening in the electronic structure due to the substantial SOC and electron correlations. \cite{Plumb2014,Kim2015,Sandilands2016,Zhou2016,Koitzsch2016} Thus, $\alpha$-RuCl$_3$ is presumed to be an SOC-assisted Mott insulator. Furthermore, a magnetic excitation continuum, which is not explained by the conventional magnon, has been reported in the paramagnetic state. \cite{Sandilands2015,Banerjee2016,Sandilands2016_2,Hirobe2017} The origin of the anomalous magnetic excitation has been discussed based on the pure Kitaev \cite{Yoshitake2016,Nasu2016,Nasu2017} and extended KH models. \cite{Winter2017,Winter2018} Coupling between the exotic magnetism and lattice degrees of freedom has also been suggested. \cite{Aoyama2017,Glamazda2017,Hasegawa2017} Recently, a field-induced phase transition to a spin-liquid state was discovered at an external field of $\sim$8 T. \cite{Leahy2017,Sears2017,Baek2017,Wolter2017,ZWang2017,Little2017,Zheng2017,Ponomaryov2017}  

Although the exotic spin-liquid state is highly expected to be the ground state of $\alpha$-RuCl$_3$ under zero external field, several phase transitions accompanied by magnetic ordering between 7 and 14 K have been reported. \cite{Kubota2015,Sears2015,Majumder2015,Banerjee2016} Synthesis of a single crystal with a unique phase transition at $\sim$14 K has been reported, where the crystal exhibited a zigzag spin structure with $2c$ magnetic superlattice modulation below 14 K. \cite{Johnson2015} However, diffuse scattering originating from the honeycomb plane (HP) stacking fault was observed in the x-ray diffraction patterns. On the other hand, another group has reported synthesis of a single crystal with no stacking fault, which exhibited a single phase transition at $\sim$7 K with $3c$ modulation. \cite{Cao2016} Those researchers also found that several phase transitions, including that at 14 K, appeared upon intentional deformation of the crystal. These results indicate that the magnetic transition temperature is sensitive to the HP stacking sequence, and that the 7 K transition is observed in only the pristine phase. 

It is important to clarify the details of the exchange interactions in $\alpha$-RuCl$_3$ in order to examine the eligibility of the KH model for elucidating the magnetic properties of this material. To this end, several theoretical studies have suggested that the direction of the ordered moment is sensitive to the signs and magnitudes of the exchange interactions. \cite{Kim2016,Winter2016,Sizyuk2016,Chaloupka2016} Further, although a neutron diffraction study has proposed possible zigzag spin structures in the 7 K phase, a large ambiguity remains with regard to the direction of the ordered magnetic moment. \cite{Cao2016} This uncertainty exists because of the scarcity of information provided by microscopic probes sensitive to the local magnetic properties. A recent muon spin rotation/relaxation ($\mu$SR) study revealed the details of the local magnetic properties in the magnetic ordered state. \cite{Lang2016} However, the polycrystalline sample exhibited only the 14 K transition, and the details of the 7 K phase as a ground state for the pristine phase continue to require clarification. 

In this paper, we report a $\mu$SR study of single-crystalline samples of $\alpha$-RuCl$_3$ to uncover the local magnetic properties of the 7 K phase. A brief description of the present $\mu$SR measurements is provided in Sec. II, which is followed by details of the data analysis of the observed $\mu$SR spectra under zero external field (ZF $\mu$SR; Sec. III A). The obtained spectra provide clear evidence that these transitions originate from a phase separation induced by the HP stacking fault, generating multiple domains with four different $T_{\rm N}$'s (Sec. III B).  In addition, the results of $\mu$SR measurements conducted under a transverse field (TF $\mu$SR) in the paramagnetic phase are analyzed in Sec. III C to refine the muon site from the magnitudes of the muon-Ru hyperfine parameters. In Sec. IV A, we discuss the muon stopping site in $\alpha$-RuCl$_3$ based on our \textit{ab initio} calculation result and the experimental hyperfine parameters. In Sec. IV B, a comparison is made between the experimental and calculated hyperfine fields at the muon site for the 7 K phase, which suggests a modulated zigzag spin structure, where the amplitude of the ordered magnetic moment is significantly reduced from the moment size expected for the orbital quenched spin-1/2 state. Finally, we discuss the sign of the Kitaev exchange interaction in $\alpha$-RuCl$_3$ from the direction of the ordered moment.

\section{Experiment}

\begin{figure}[h]
\includegraphics[width=7.5cm,clip]{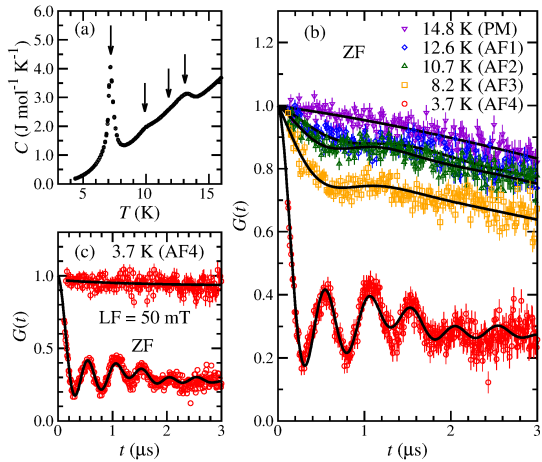}
\caption{\label{fig:tspect} (a) Temperature dependence of specific heat of a single crystal used in the present $\mu$SR experiment. Arrows represent anomalies indicating the four magnetic ordering transitions at $T_{\rm N1}$--$T_{\rm N4}$. (b) Zero-field (ZF) $\mu$SR time spectra $G(t)$ measured at selected temperatures. The solid curves represent the results of least-squares fitting using Eq. (\ref{eq:Asym}). (c) $\mu$SR time spectra at 3.7 K measured under ZF and longitudinal field (LF) of 50 mT.}
\end{figure}

Conventional ZF $\mu$SR experiments were conducted using the Advanced Research Targeted Experimental Muon Instrument at S-line (ARTEMIS) spectrometer equipped on the S1 beamline of the Japan Proton Accelerator Research Complex (J-PARC). We used several thin platelike $\alpha$-RuCl$_3$ single crystals. Figure \ref{fig:tspect}(a) shows the temperature $T$ dependence of the specific heat $C$ for one of the single crystals used in the present $\mu$SR experiment. From the data, we found that the four phase transitions were observed at $T_{\rm N1}$ = 13.1(2) K, $T_{\rm N2}$ = 11.8(2) K, $T_{\rm N3}$ = 9.9(2) K, and $T_{\rm N4}$ = 7.2(1) K, where $T_{\rm N4}$ corresponds to the transition temperature of the AFM 7 K phase with the least stacking fault. \cite{Cao2016} Details of the crystal growth were reported elsewhere. \cite{Kubota2015}   The crystal surface was parallel to the HP, while the initial muon spin polarization was perpendicular to the HP. The muon spin depolarization function $G(t)$ was obtained from the decay-positron asymmetry.   

We also performed TF $\mu$SR measurements using the Nu-Time spectrometer installed on the TRIUMF M15 beamline, where the external field $B_0$ was perpendicular to the initial muon spin polarization.  $B_0$ was applied either parallel or perpendicular to the HP.  A sample with dimensions of $8.1 \times 7.6 \times 0.9$ mm$^3$ was used for the measurements with $B_0 \perp$ HP. For those with $B_0 \parallel$ HP, we employed four plate-like crystals with typical dimensions of $6.5 \times 2.6 \times 1.6$ mm$^3$ stacked along the direction perpendicular to the HP.    The precise magnitudes of $B_0$ were determined to be $B_0 = 5.99952(2)$ and $5.99956(5)$ T for $B_0 \perp$ HP and $B_0 \parallel$ HP, respectively, where the calibration was performed based on the  precession frequency of muons stopped in a  ``muon-veto"  scintillator (made of CaCO$_3$) mounted beneath the sample in the cryostat. The positron signals from the sample and the CaCO$_3$ were separated by sorting the positron events using the signal from the muon-veto scintillator.

\section{Results and analysis}

\subsection{Zero-field $\mu$SR spectra}

Implanted muons probe magnetic order through the precession of the muon magnetic moment under the internal field $B_{\rm int}$ associated with the spontaneous magnetization. As shown in Fig.~\ref{fig:tspect}(b), we detected the development of oscillating signals in the ZF $\mu$SR spectra at lower temperatures, which is a clear indication of the appearance of the quasistatic magnetic order. Note that we have abbreviated the $T$ ranges above $T_{\rm N1}$, $T_{\rm N2} < T \leq T_{\rm N1}$, $T_{\rm N3} < T \leq T_{\rm N2}$, $T_{\rm N4} < T \leq T_{\rm N3}$, and $T \leq T_{\rm N4}$ as PM, AF1, AF2, AF3, and AF4, respectively.  

\begin{figure}[t]
\includegraphics[width=7.5cm,clip]{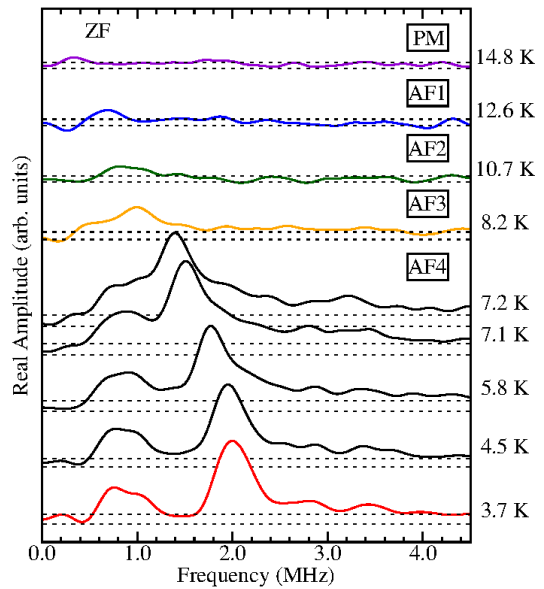}
\caption{\label{fig:fft} Fast Fourier transform of ZF $\mu$SR spectra at selected temperatures. The dotted lines represent the upper and lower limits of the estimated noise level. The line colors correspond to those for the time spectra shown in Fig. \ref{fig:tspect}(b). }
\end{figure}

Considering the fast Fourier transform (FFT) of the ZF $\mu$SR spectra shown in Fig.~\ref{fig:fft}, we made the least-squares curve-fit analysis of $G(t)$ using the following recursive function,
\begin{eqnarray}
	G(t) = && (1 - f_{\rm m}) G_{\rm KT}(t) e^{-\lambda_{\rm p} t} + f \cos(2 \pi \nu t) e^{-\lambda t} \nonumber \\
		 && + f_{\rm B} j_0 (2 \pi \nu_{\rm B}t) e^{-\lambda_{\rm B} t} + f_{\rm c} e^{-\lambda_{\rm c} t}, 
\label{eq:Asym}
\end{eqnarray}
where the first term is the paramagnetic component described by a product of exponential damping with a rate $\lambda_{\rm p}$ due to dynamical field from the fluctuating electron spins and the Kubo-Toyabe function \cite{KT}  $G_{\rm KT}(t)$ for the depolarization due to random local fields from nuclear magnetic moments. The second and third terms represent precession signals from the AFM phases, where $j_0$ represents the Bessel function, $f$ and $f_{\rm B}$ are the fractional yields, $\nu$ and $\nu_{\rm B}$  ($2 \pi \nu = \gamma_{\mu} B_{{\rm int}}$) are precession frequencies with  $\gamma_{\mu} = 2 \pi \times 135.5388$ MHz/T, $B_{\rm int}$ is the internal field at the muon site proportional to sublattice magnetization, $\lambda$ and  $\lambda_{\rm B}$ are the depolarization rates of the respective signals monitoring the distribution and/or fluctuation of $B_{\rm int}$. The last term represents the non precessing component corresponding to a fraction of implanted muons $f_{\rm c}$ subjected to $B_{\rm int}$ parallel to the initial muon spin polarization, thereby showing weak longitudinal depolarization with $\lambda_{\rm c}$.  The total signal fraction of the magnetically ordered phases is denoted by $f_{\rm m}$ in the first term, where $f_{\rm m} = f + f_{\rm B} + f_{\rm c}$. The curve fit using Eq.~(\ref{eq:Asym}) yielded satisfactory conversion for every time spectra with an averaged normalized $\overline{\chi^2} = 1.3(2)$. Here, the use of the Bessel function in Eq.~(\ref{eq:Asym}) is justified by the previous neutron diffraction study that suggested the possibility of incommensurate magnetic modulation due to pseudo-three-layer HP stacking, although the incommensurate modulation period was too long to be detected by the neutron diffraction experiment. \cite{Cao2016}  We also note that a preliminary analysis using  Eq.~(\ref{eq:Asym}) with $\cos(2\pi\nu_{\rm B}t)$ substituted for $j_0(2\pi\nu_{\rm B}t)$ yielded strongly $T$ dependent $f_{\rm B}$ and $f$, which is unlikely for the fully developed magnetic phases, thus supporting the use of the Bessel function.

\begin{figure}[h]
\includegraphics[width=7.5cm,clip]{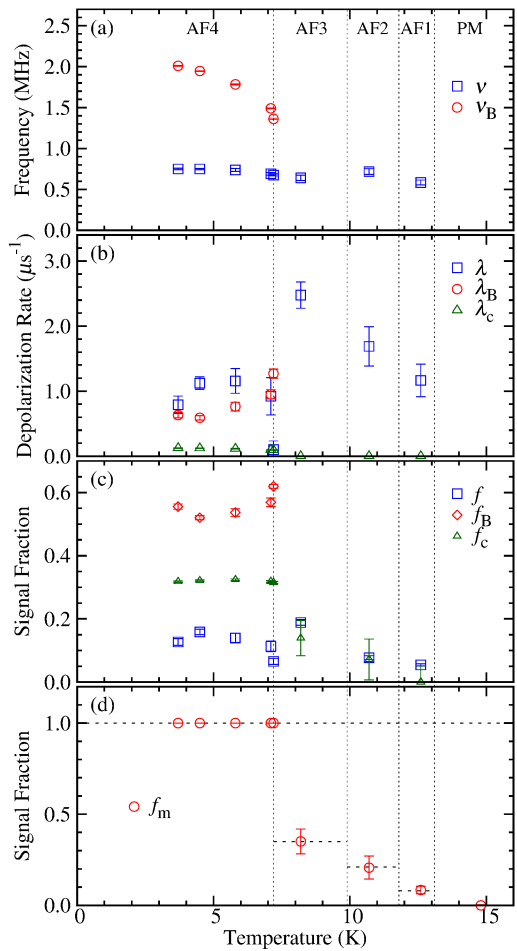}
\caption{\label{fig:ZF_Tdep} Temperature dependence of (a) precession frequencies, (b) depolarization rate, (c) fractional yield, and (d) the sum of fractional yield for magnetically ordered phases ($f_{\rm m} = f + f_{\rm B} + f_0$). The dashed lines in (d) are guides for the eye. }
\end{figure}     

\subsection{Phase separation}

Since the local field $B_{\rm int}$ is mainly determined by transferred hyperfine field and magnetic dipolar field from ordered magnetic moments within a distance of several nanometers around the muon, $\mu$SR is inherently sensitive to spatial phase separation.  Provided that a sample (or a domain within the sample) exhibits spatially uniform and multistage phase transitions at several critical temperatures, these transitions are observed as a stepwise change in $B_{\rm int}$ and in the corresponding precession frequency at each transition temperature. Meanwhile, if a spatial phase separation occurs, allowing different magnetic structures with different transition temperatures and $B_{\rm int}$'s to coexist as macroscopic domains, multiple precession signals develop independently from zero frequency at each transition temperature, with their relative yields corresponding to their respective volumetric fractions.  

As summarized in Fig. \ref{fig:ZF_Tdep}(a), we find one precession signal with a frequency $\nu$ below $T_{\rm N1}$, which is in line with a previous report. \cite{Lang2016} Upon cooling, another precession signal (previously unobserved) appears below $T_{\rm N4}$ that is reproduced by the Bessel function with a frequency $\nu_{\rm B}$, while no anomaly is found for $\nu$ around $T_{\rm N4}$. The relatively small $\lambda_{\rm B}$ [$\ll 2\pi\nu_{\rm B}$; see Fig.~\ref{fig:ZF_Tdep}(b)] implies that these multiple phases have a macroscopic domain size that far exceeds the effective range of the magnetic dipole field exerted on muons from Ru moments. These observations strongly suggest that a phase separation occurs within the sample, and that each magnetic phase has different critical temperatures. The sinusoidal signal is reasonably attributed to the 14 K phase, whereas the Bessel term corresponds to the precession signal for the pristine 7 K phase. Note that $\mu$SR spectra for the short-range ordered state are expected to exhibit Gaussian or exponential-like damping without oscillation. \cite{Kanigel2002,Dally2014} The macroscopic domain size is also supported by the presence of a well-defined thermodynamic phase inferred from a clear phase transition at $T_{\rm N4}$. \cite{Kubota2015} 

The temperature dependence of the total signal fraction of the magnetic ordered states $f_{\rm m}$ is plotted in Fig. \ref{fig:ZF_Tdep}(d). If the four magnetic transitions occur successively in a single-phase sample,  $f_{\rm m}$ should exhibit a steep change to unity immediately below $T_{\rm N1}$. However,  it is clear in Fig. \ref{fig:ZF_Tdep}(d) that $f_{\rm m}$ actually exhibits a gradual increase with decreasing $T$ above $T_{\rm N4}$ to saturate below $T_{\rm N4}$. This further supports the occurrence of phase separation in this sample. The volumetric fraction of the 7 K phase is roughly estimated to be $\sim$65 \%. We stress that the 7 K phase has the largest volume in our sample. 

According to a previous $\mu$SR study, the development of 14 K phase accompanying a precession signal of $\sim$1 MHz is followed by an additional signal with a large frequency distribution  below $\sim$11 K. \cite{Lang2016} While the small-volume fraction of the 14 K phase for the present $\mu$SR data does not allow a curve fit with these multiple signals, the enhanced $\lambda$ in the relevant temperature region can be interpreted as being due to the occurrence of unresolved precession signals. The sudden decrease of $\lambda$ below $T_{\rm N4}$ is explained by the correlation of parameters in the curve fit, where part of the depolarization corresponding to the 14 K phase was represented by the Bessel function. Considering the small fraction of the additional 14 K phase signal appeared below $\sim$11 K ($\sim$5\%--10\%), the correlation would not affect the primary feature of the deduced result for the 7 K phase.

\subsection{Hyperfine fields in the paramagnetic phase}

\begin{figure}[h]
\includegraphics[width=7cm,clip]{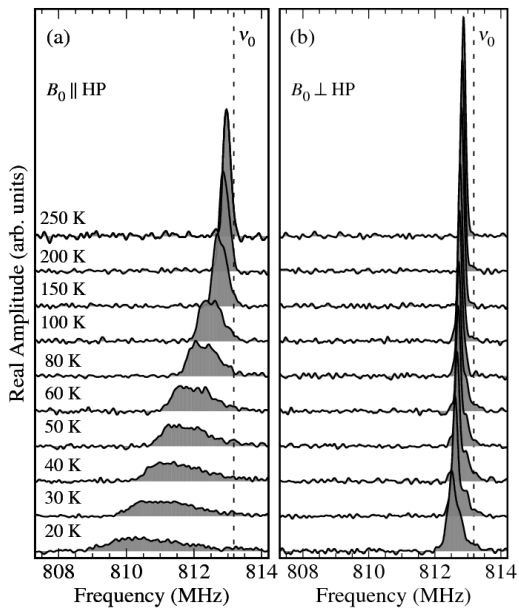}
\caption{\label{fig:fft_TF} Temperature variations of fast Fourier transformed $\mu$SR spectra measured under transverse field. The external field $B_0$ was applied (a) parallel and (b) perpendicular to the honeycomb plane (HP), where $B_0 = 5.99956(5)$ T for $B_0 \parallel$ HP and $B_0 = 5.99952(2)$ T for $B_0 \perp$ HP. The dashed lines correspond to the reference frequency. }
\end{figure}

The frequency shifts deduced from the TF $\mu$SR spectra provide the magnitudes of the muon-Ru hyperfine parameters and their anisotropies, which constitute information valuable for identifying the muon site. Figure \ref{fig:fft_TF}(a) shows the FFT of TF $\mu$SR spectra obtained for $B_0$ applied parallel to the HP, where we find two peaks with similar amplitudes. These peaks exhibit shifts to lower frequencies and broadening with decreasing $T$. The corresponding TF $\mu$SR spectra were analyzed in the time domain via curve fitting using two cosine functions with the Gaussian envelope
\begin{equation}  
\begin{split}
G_{\parallel}(t) = & f_{\parallel 1} \cos(2 \pi \nu_{\parallel 1} t + \theta_0) \exp[-(\lambda_{\parallel 1} t)^2] \\
& + f_{\parallel 2} \cos(2 \pi \nu_{\parallel 2} t + \theta_0) \exp[-(\lambda_{\parallel 2} t)^2].
\label{eq:TF_para}
\end{split}
\end{equation}
On the other hand, the FFT spectra for $B_0 \perp$ HP exhibit an almost $T$-independent feature with a single sharp line, as shown in Fig. \ref{fig:fft_TF}(b). These spectra were analyzed using 
\begin{equation} 
\begin{split}
G_{\perp}(t) = & f_{\perp} \cos(2 \pi \nu_{\perp} + \theta_0) \exp[-(\lambda_{\perp} t)^{\beta}] \\ 
& + (1 - f_{\perp}) \cos(2 \pi \nu_{\rm ex} + \theta_0) \exp(-\lambda_{\rm ex} t),
\label{eq:TF_perp}
\end{split}
\end{equation}
where $\theta_0$ is the initial phase and $\beta$ is the stretched exponent. The second term was adopted to describe the additional small peak, which we tentatively attributed to a background signal of unknown origin.  

As will be discussed in Sec.~IV A in more detail, the observed dependence of the frequency shift on the $B_0$ direction is consistent with the assumption that $\nu_{\parallel 1}$, $\nu_{\parallel 2}$, and $\nu_{\perp}$  originate from a single muon site with the corresponding anisotropy in the local susceptibility.  More specifically, the observed spectrum for $B_0 \parallel$ HP is explained as a convolution of spectra with $B_0$ applied along the various directions parallel to the HP. This explanation can be made because we used several crystals for the corresponding measurements, where each crystal had domains due to the monoclinic structure. Thus, the two observed peaks can be attributed to an approximately two-dimensional powder pattern for a single muon site. 

We have already shown that a phase separation occurs in the present sample. However, we stress that the local magnetic property in the paramagnetic state is not affected by the stacking fault, because the magnetic susceptibility above $\sim$30 K is almost independent of the degree of complexity in the  magnetically ordered phases below $\sim$14 K.\cite{Tanaka}  Thus, we can presume without much uncertainty that the phase separation is irrelevant to the splitting of TF $\mu$SR frequency spectra in the paramagnetic state. 

\begin{figure}[h]
\includegraphics[width=7.5cm,clip]{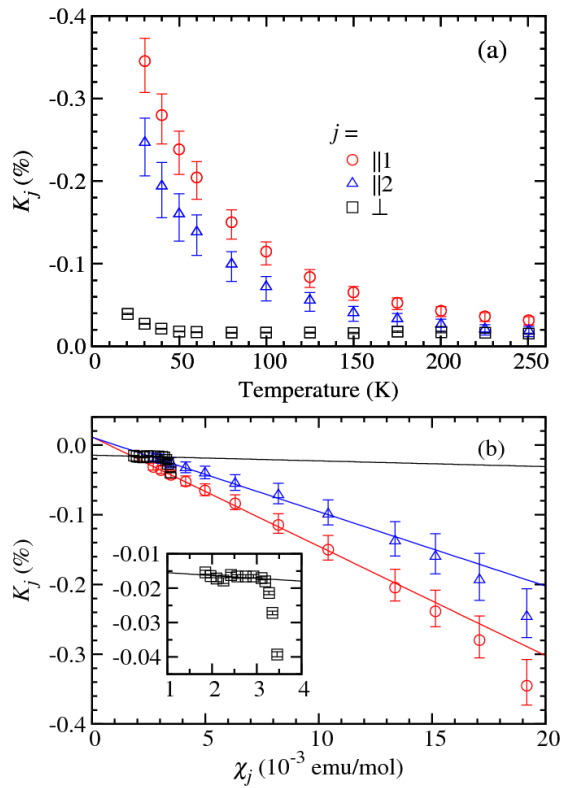}
\caption{\label{fig:shift} (a) Temperature dependence of the frequency shift, where the circles, triangles, and squares correspond to $K_{\parallel 1}$, $K_{\parallel 2}$, and $K_{\perp}$, respectively. The vertical axis is reversed for clarity. (b) $K_j$ plotted against magnetic susceptibility $\chi_j$, with the temperature being an internal variable. The solid lines were obtained via curve fitting using Eq. (\ref{eq:K_chi}) for the data above 50 K. The inset shows an enlarged view of the $K_{\perp}$ versus $\chi_{\perp}$ plot. }
\end{figure}

The frequency shift ($K_j$, $j =$ $\parallel$1, $\parallel$2, and $\perp$) is induced by additional hyperfine fields at the muon site due to polarized Ru electron spins. We deduced $K_j$ by subtracting the contributions from the Lorentz field and demagnetization, such that \cite{Yasuoka1978}
\begin{equation}
K_j = \frac{\nu_j - \nu_0}{\nu_0} - 4 \pi (\frac{1}{3} - N_j) \rho \chi_j,
\label{eq:shift}
\end{equation}
where $2 \pi \nu_0 = \gamma_{\mu} B_0$, $N_j$ ($N_{\parallel 1} = N_{\parallel 2} \simeq 0.28$--0.41, $N_{\perp} \simeq 0.84$--0.85) is the demagnetization factor,\cite{demag} $\rho$ ($= 0.01884$ mol/cm$^3$) is the molar density of $\alpha$-RuCl$_3$, and $\chi_j$ is the molar susceptibility.  The obtained values for $K_{\parallel 1}$, $K_{\parallel 2}$, and $K_{\perp}$ versus temperature are shown in Fig.~\ref{fig:shift}(a). Note that the vertical axis is reversed to highlight the increasing behavior of the local magnetic susceptibility with decreasing $T$.

\begin{table*}
\caption{\label{tab:A} Potential energies $U$ at possible muon sites measured from that for the grand minimum, along with the Wyckoff position, relative position, point symmetry for the $C2/m$ structure, and calculated dipole contribution for the hyperfine parameters for the four expected muon sites $\mu(k)$ ($k$ = 1, 2, 3, and 4).  $A_{\rm iso} = \frac{1}{3}(A_{\perp} + A_{\parallel 1} + A_{\parallel 2})$, $A_{\rm ax} = \frac{1}{6}[2A_{\perp} - (A_{\parallel 1} + A_{\parallel 2})]$, and $A_{\rm asym} = \frac{1}{2} (A_{\parallel 1} - A_{\parallel 2})$, corresponding to the isotropic, axial, and asymmetric components of the hyperfine coupling tensor, respectively. The experimentally evaluated results are shown in the last row. }
\begin{ruledtabular}
\begin{tabular}{ccccccccc}
Muon site & $U$ (eV) & Wyckoff position & Relative position & Point symmetry & $A_{\rm iso}$ (mT/$\mu_{\rm B}$) & $A_{\rm ax}$ (mT/$\mu_{\rm B}$) & $A_{\rm asym}$ (mT/$\mu_{\rm B}$) \\
\hline
$\mu(1)$ & 0.00 & $2d$ &  (0, 1/2, 1/2) & $2/m$ & 0 & -56 &   35 \\
$\mu(2)$ & 0.23 & $4i$ & (1/3, 0, 0.1) &  $m$  & 0 &  35 &   -12 \\
$\mu(3)$ & 0.32 & $4g$ &  (0, 0.28, 0)  &  $2$  & 0 &  54 & -16 \\
$\mu(4)$ & 0.38 & $2a$ &  (0, 0, 0) & $2/m$ & 0 &  46 & -31 \\
Experiment &   &   &   &   & -51(4) & 23(1) & -14(1) \\
\end{tabular}
\end{ruledtabular}
\end{table*}

It is highly likely that the in-plane magnetic anisotropy is weak in this compound, as can be inferred from the relatively small difference between $K_{\parallel 1}$ and $K_{\parallel 2}$. Thus, we assume $\chi_{\parallel 1} \simeq \chi_{\parallel 2} = \chi_{\perp}$. In Fig. \ref{fig:shift}(b), $K_j$ are plotted against $\chi_j$ with the temperature as an internal variable, where $\chi_{\parallel}$ and $\chi_{\perp}$ are quoted from an earlier study.\cite{Kubota2015} It is generally expected that $K_j$ can be described by 
\begin{equation}
K_{j}(T) = K_0 + \frac{A_{j}}{N_{\rm A} \mu_{\rm B}} \chi_{j}(T),
\label{eq:K_chi}
\end{equation}
where $K_0$ refers to the $T$-independent contribution, $A_{j}$ are the hyperfine parameters between the muon spin and the electron spins, $N_{\rm A}$ is Avogadro's number, and $\mu_{\rm B}$ is the Bohr magneton. As $\alpha$-RuCl$_3$ exhibits anisotropic magnetic susceptibility ($\chi_{\parallel}\neq\chi_{\perp}$), $K_{\parallel 1}$ and $K_{\parallel 2}$ are plotted against $\chi_{\parallel}$, whereas $\chi_{\perp}$ is used for the $K_{\perp}$ case. As shown in Fig. \ref{fig:shift}(b), we found a linear relationship between $K_j$ and $\chi_j$ above 50 K and obtained $A_{\parallel 1} = -88(2)$ mT/$\mu_{\rm B}$, $A_{\parallel 2} = -60(2)$ mT/$\mu_{\rm B}$, and $A_{\perp} = -5(3)$ mT/$\mu_{\rm B}$ from the slopes. To characterize the hyperfine coupling tensor, we defined the isotropic, axial, and asymmetric components, $A_{\rm iso} = \frac{1}{3} (A_{\perp} + A_{\parallel 1} + A_{\parallel 2}) = -51(4)$ mT/$\mu_{\rm B}$, $A_{\rm ax} = \frac{1}{6} [2 A_{\perp} - (A_{\parallel 1} + A_{\parallel 2})] = 23(1)$ mT/$\mu_{\rm B}$, and $A_{\rm asym} = \frac{1}{2}(A_{\parallel 1} - A_{\parallel 2}) = -14(1)$ mT/$\mu_{\rm B}$ mT/$\mu_{\rm B}$. These values are summarized in Table \ref{tab:A}.

\begin{figure}[h]
\includegraphics[width=7.0cm,clip]{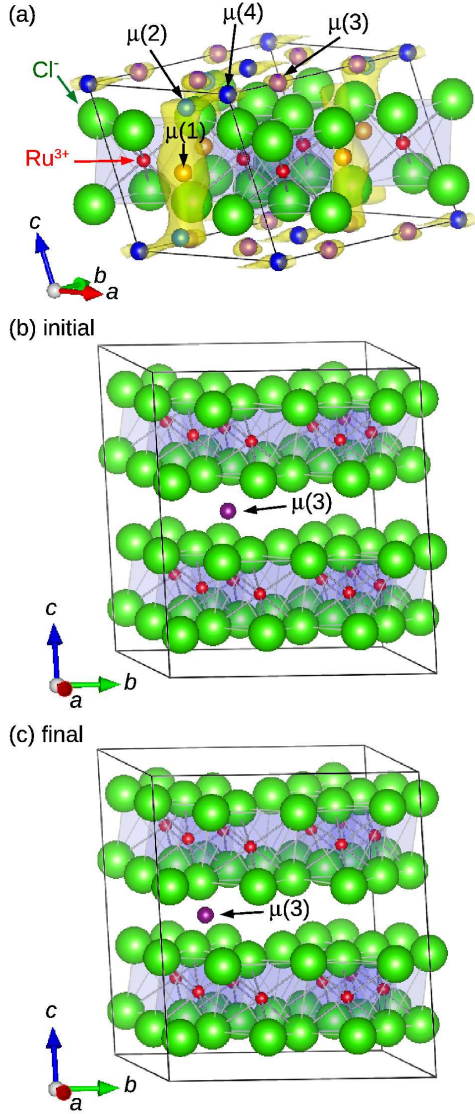}
\caption{\label{fig:musite} (a) Schematic illustration of crystal structure for $\alpha$-RuCl$_3$. The four possible muon sites $\mu(k)$ ($k$ = 1, 2, 3, and 4), which are obtained with the Hartree potential calculation, are highlighted by orange, light blue, purple, and dark blue balls, respectively. The yellow hatching indicates regions in which the Hartree potential for a positive muon $U$ measured from the grand minimum is lower than 0.40 eV. (b) The initial $2 \times 1 \times 2$ supercell structure with a positive muon (proton) placed at the $\mu(3)$ site used in our relaxed structure calculation. (c) The obtained relaxed crystal structure for the calculation. These plots were drawn using VESTA. \cite{VESTA}}
\end{figure}

\section{Discussion}

\subsection{Muon site}

Our Hartree potential calculation suggests four different muon sites corresponding to local potential minima  ($U \leq$ 0.4 eV, where $U$ is the potential energy measured from the minimum), which are illustrated by yellow hatched areas in Fig. \ref{fig:musite}(a). The calculation was done using the Vienna Ab initio Simulation Package (VASP) \cite{VASP} which can calculate the electron density distribution in the atomic positions and pseudo potential based on the density functional theory (DFT) with the generalized gradient approximation (GGA). We labeled these candidate sites $\mu(k)$, with $k$ = 1 to 4, for which the magnitudes of $U$, along with the Wyckoff positions, relative positions, and point symmetries, are summarized in Table \ref{tab:A}. Note that a similar result has been reported in Ref. [\onlinecite{Lang2016}] where the origin of the unit cell is shifted from ours by $(0,1/2,1/2)$. 

In insulators, it has been suggested that a positive muon modifies the atomic positions of the surrounding ions. \cite{Bernardini2013,Moller2013} To further refine the potential minima, we  calculated the relaxed crystal structure with \textit{ab initio} calculations. The calculation was made with using the Open source package for Materials eXplorer (OpenMX) code, which is based on the DFT+GGA and norm-conserving pseudo potential method. \cite{OpneMX} We used a charged $2 \times 1 \times 2$ supercell with a proton, which mimics a positive muon, placed at each of the expected muon site candidates as the initial structure. Schematic plot of the initial and relaxed structures for the $\mu$(3) case are illustrated in Figs. \ref{fig:musite}(b) and \ref{fig:musite}(c). All the $\mu(k)$ sites showed 0.8--1.3 {\AA} displacement from the initial positions except for the $\mu(4)$ site. 

Furthermore, we compared the experimental hyperfine parameters with those calculated for the candidate sites. Generally, the hyperfine field in a non metallic compound is primarily due to the isotropic transferred hyperfine interaction and anisotropic magnetic dipolar interaction between the muon and local electronic moments. Thus, we evaluated the anisotropic hyperfine parameters by calculating the classical dipole field from the 1$\mu_{\rm B}$ spins at the Ru sites, within a sphere with a 5-nm radius. We used the Ru sites in the relaxed $2 \times 1 \times 2$ structures for the nearest 16 Ru sites for the calculation. The calculated $A_{\rm iso}$, $A_{\rm ax}$, and $A_{\rm asym}$ are also listed in Table \ref{tab:A}. It must be noted that the dipolar hyperfine field is purely anisotropic, yielding $A_{\rm iso}=0$.  Therefore, the observed finite value for $A_{\rm iso}$ may be attributed to the transferred hyperfine interaction due to unpaired electron spin density at the muon site. \cite{Yaouanc} In the following, we focus on the comparison between $A_{\rm ax}$ and $A_{\rm asym}$ by assuming that the anisotropic components are dominated by the dipole interaction. 

The $\mu(1)$ site is located near the center of the Ru hexagon in the HP, which is parallel to the $ab$ plane in Fig. \ref{fig:musite}(a). For the $\mu(1)$ site, a large negative $A_{\rm ax}$ is given by our dipole field calculation, as shown in Table \ref{tab:A}, indicating a large positive frequency shift for $B_0 \parallel$ HP. However, we obtained large negative values for $K_{\parallel 1}$ and $K_{\parallel 2}$, and the experimental $A_{\rm ax}$ was positive. Thus, we can reject the $\mu(1)$ site, although it is at the potential minimum. All the other muon site candidates are located in the inter-HP sites, and positive $A_{\rm ax}$ and negative $A_{\rm asym}$ are expected. Although we could not find a muon site candidate showing perfect agreement between the experimental and calculated results, the $\mu(2)$, $\mu(3)$, and $\mu(4)$ sites are favored.

\begin{figure}
\includegraphics[width=7.5cm,clip]{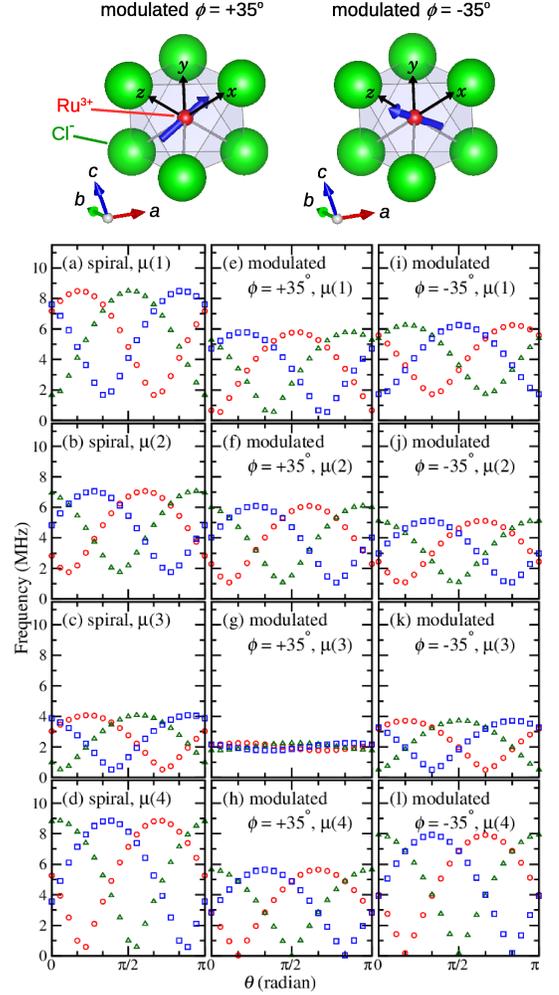}
\caption{\label{fig:simulation} Schematic illustration of two canting directions for the modulated zigzag spin structure depicted on a local rectangular coordination. The red and green balls correspond to the Ru$^{3+}$ and Cl$^-$ ions, respectively. The blue arrows are the ordered magnetic moments (drawn using VESTA\cite{VESTA}). Calculated precession frequency $\mu(k)$ ($k$ = 1--4) sites for (a)--(d) the spiral zigzag spin structure, and for the modulated zigzag spin structure with (e)--(h) a canted angle $\phi = +35^{\circ}$ and (i)--(l) $\phi = -35^{\circ}$. \cite{Cao2016} The calculated frequencies are plotted against the initial phase of the $3c$ modulation of the ordered moment. The circles, triangles, and squares correspond to the precession frequencies for the respective $\mu(k)$ positions in the $a \times b \times 3c$ magnetic unit cell, where the internal fields differ from each other. }
\end{figure}

\subsection{Spin structure of the 7 K phase}

A recent neutron diffraction study proposed two types of possible spin structures for the ground state, i.e., spiral and modulated zig-zag spin structures. \cite{Cao2016} Both of these structures have an ordered magnetic moment aligned in a zig-zag manner within each honeycomb layer, accompanied by $3c$ magnetic superlattice modulation (see Fig.~5 in Ref. [\onlinecite{Cao2016}]). It also pointed out the possibility of an incommensurate magnetic modulation, where the modulation period is too long to be detected by the neutron diffraction experiment since the HP stacking period slightly deviates from $3c$. \cite{Cao2016} We consider that the observed Bessel term in our ZF $\mu$SR spectra is related to the possible incommensurate modulation. However, we discuss the spin structure by comparing our experimental results and the spin structures with the 3$c$ modulation because the accurate periodicity of the incommensurate modulation has not been determined at present. 

For the spiral zigzag spin structure, the ordered moment rotates in the $ca$ plane from layer to layer. The ordered moment of the $n$th layer ${\bf m}_n$ is described as ${\bf m}_n = m_0 [ \cos(\theta \pm \frac{2}{3} \pi n), 0, \sin(\theta \pm \frac{2}{3} \pi n) ]$ in the rectangular $ab$$\perp$ coordinate system, where $m_0$ is the magnitude of the ordered moment and $\theta$ is the initial phase. It is estimated that $m_0 = 0.45(5) \mu_{\rm B}$. On the other hand, in the modulated zigzag spin structure, the direction of the ordered magnetic moment is fixed, and the amplitude modulates from layer to layer. The magnetic moment lies in the $ca$ plane, tilted from the $a$ axis with angle $\phi =+35^{\circ}$ or $-35^{\circ}$, as schematically illustrated in Fig.~\ref{fig:simulation}. The ordered magnetic moment lies approximately within the $xy$ plane of the RuCl$_6$ octahedron for $\phi = +35^{\circ}$, whereas it is oriented in the $z$ direction for $\phi = -35^{\circ}$. The magnitude of the ordered moment at the $n$-th layer is $m_n = m_0 \sin(\theta \pm \frac{2}{3}\pi n)$, with $m_0 = 0.60(5) \mu_{\rm B}$.    

In order to assess these two types of spin structures in view of our experimental results, we calculated $B_{\rm int}$ at the four candidate muon sites, where $B_{\rm int}$ was evaluated as a function of $\theta$ (the initial phase of the $3c$ modulation) by taking the sum of classical dipole fields from the ordered magnetic moments within a sphere with a 5-nm radius around the $\mu(k)$ site. As shown in Sec. IV A, we used the 16 Ru sites in the $2 \times 1 \times 2$ relaxed structure as the nearest 16 Ru sites in the sphere. Figure \ref{fig:simulation} shows the $\theta$ dependence of the calculated muon precession frequencies at each $\mu(k)$, and one can observe several precession frequencies at a given $\theta$. This is because there are several $\mu(k)$ positions in the $a \times b \times 3c$ magnetic unit cell; for instance, there are six $\mu(1)$ positions in the magnetic unit cell, and the internal field can differ at each position. If the incommensurate modulation period is sufficiently long, the calculated precession frequency corresponds to the peaked frequency $\nu_{\rm B}$ for the Bessel term. Interestingly, these frequencies approximately degenerate to a single value at $\sim$2 MHz for the $\mu(3)$ site with the $\phi = +35^{\circ}$ modulated zig-zag spin structure [see Fig.~\ref{fig:simulation}(g)], which is in agreement with our experimental observation of a single precession signal at $\sim$2 MHz below $T_{\rm N4}$.  Accordingly, it is suggested that the implanted muon stops at the $\mu(3)$ site and that the $\phi = +35^{\circ}$ modulated zig-zag spin structure appears below $T_{\rm N4}$, while $\theta$ remains unknown. Further calculation confirmed that such degeneracy appeared within a $\phi$ range of only $+30^{\circ} \lesssim \phi \lesssim +45^{\circ}$. 

The Ru moment size ($\sim0.6 \mu_{\rm B}$) inferred from the previous neutron diffraction study \cite{Cao2016} is significantly smaller than the saturated moment 1$\mu_{\rm B}$ for the orbital-quenched spin-1/2 system, suggesting the existence of a large orbital contribution to the microscopic magnetic properties of $\alpha$-RuCl$_3$. As a possible origin of the small ordered magnetic moment, the presence of residual magnetic fluctuation is suggested. \cite{Cao2016,Sandilands2015} Meanwhile, the present $\mu$SR result including the spectra at 2 K under various longitudinal fields  [applied along the initial muon spin polarization, shown in the Fig.~\ref{fig:tspect}(c)] indicates that $B_{\rm int}$ is predominantly determined by the reduced quasistatic Ru moments. Thus, we speculate that the origin of the moment shrinkage may be due to magnetic excitation that accompanies fast fluctuation modes perpendicular to the magnetization axis.
 
According to a recent theoretical study, the direction of the ordered magnetic moment in the magnetic ordered state is a sensitive criterion for evaluating the exchange interactions. Chaloupka and Khaliullin clarified that the zigzag magnetic ordered state with the ordered magnetic moment lying in the $xy$ plane is stabilized for the extended KH model in which the third-nearest-neighbor exchange interaction $J_3$ and ferromagnetic (FM) $K$ are included, whereas the conventional KH model with AFM $K$ and FM $J$ predicted the zigzag spin structure with the ordered magnetic moment pointing along the $z$ axis. \cite{Chaloupka2016} The present $\mu$SR result is clearly in favor of the former scenario.  

Recently, the extended KH model with FM $K$ and AFM $J$ received further theoretical support from the combined quantum chemistry electronic-structure calculation and exact-diagonalization method. \cite{Yadav2016} Winter \textit{et al.} also proposed an extended KH model with the FM Kitaev term being an effective spin Hamiltonian by using the exact-diagonalization method, and they also reinterpreted the inelastic neutron scattering results in Ref. [\onlinecite{Banerjee2016,Banerjee2017}] based on their effective spin Hamiltonian with the FM Kitaev interaction. \cite{Winter2016,Winter2017} Recent inelastic neutron scattering studies also proposed the FM Kitaev term. \cite{Ran2017,Do2017,Banerjee2017FI} Wang \textit{et al.} also explained the neutron scattering experiment results based on the spin-wave theory for their $K$-$\Gamma$ model with FM $K$. \cite{WWang2017}     

As a remaining issue, we discuss the effect of the transferred hyperfine interaction $A_{\rm tr}$ in the magnetic ordered state, which was not considered in the above internal field calculation. As mentioned in Sec. III C, the presence of isotropic hyperfine coupling $A_{\rm iso}$ points to the possibility of transferred hyperfine-type interaction for its origin, $A_{\rm iso} \sim A_{\rm tr}$. However, considering that it might be short range and exerted from several neighboring magnetic ions, we conclude that such a hyperfine field would be canceled out for the $\mu(1)$, $\mu(2)$, and $\mu(4)$ sites due to the local symmetry for the proposed spin structures. On the other hand, finite transfered hyperfine field might be expected at the $\mu(3)$ site because the distances between the $\mu(3)$ site and the neighboring magnetic ions are different from each other and the vector sum of the magnetic moments at the neighboring magnetic ions remains finite. In this regard, we need quantitative estimation of the transferred hyperfine field to examine the consistency of our result with the spin structure presumed for the 7 K phase, which is beyond the scope of the present study. 

We also note that the calculated precession frequency shown in Fig. \ref{fig:simulation} is sensitive to the position of the muon site. For the muon sites suggested by the VASP calculation without optimization for structural relaxation, the estimated muon precession frequencies at the site corresponding to $\mu(3)$ degenerated to $\sim$2.7 MHz for the $\phi = -35^{\circ}$ modulated zigzag spin structure.  This is apparently in favor of an AFM $K$, in contrast to the conclusion drawn above, although the calculated frequency is slightly off the observed value of $\sim$2 MHz. 
Thus, more reliable information about the muon site is required to come to a definite conclusion based solely on the muon experiment.

\section{Summary}

We performed $\mu$SR measurements on single-crystalline samples of the $\alpha$-RuCl$_3$ honeycomb magnet, which is a candidate compound for the Kitaev-Heisenberg model. The samples exhibited magnetic phase transitions at $T_{\rm N1}$=13.1(2) K, $T_{\rm N2}$=11.8(2) K, $T_{\rm N3}$=9.9(2) K, and $T_{\rm N4}$=7.2(1) K, where the AFM 7 K phase corresponds to that with the least stacking fault. Our $\mu$SR result indicates that the successive phase transitions originate from spatial phase separation, most likely due to the honeycomb plane stacking fault reported by Cao $\textit{et al}$. \cite{Cao2016} The frequency shift inferred from TF $\mu$SR measurements in the paramagnetic state strongly suggests that the implanted muons are stopped in the interlayer site between the honeycomb planes. A detailed assessment of the present $\mu$SR result combined with the DFT calculation for muon sites suggests that the $\phi = +35^{\circ}$ modulated zigzag spin structure, which has been proposed to be one of the possible spin structures based on a neutron diffraction study, is the most plausible magnetic structure for the 7 K phase. \cite{Cao2016}

\begin{acknowledgments}
We thank H. Lee, T. Ishibashi, T. Yuasa, and the staff of J-PARC MUSE and TRIUMF for their technical support. This work was partially supported by the Condensed Matter Research Center, Institute of Materials Structure Science, KEK, the KEK Large Scale Simulation Program (Nos. 15/16-07 and 16/17-18) and Izumi Science and Technology Foundation (Grant No. H28-J-135).  
\end{acknowledgments}


\begin{thebibliography}{99}

\bibitem{Kitaev2006} A. Kitaev, Ann. Phys. (N.Y.) \textbf{321}, 2 (2006).

\bibitem{Jackeli2009} G. Jackeli and G. Khaliullin, Phys. Rev. Lett. \textbf{102}, 017205 (2009).
\bibitem{Chaloupka2010} J. Chaloupka, G. Jackeli, and G. Khaliullin, Phys. Rev. Lett. \textbf{105}, 027204 (2010).
\bibitem{Rau2014} J. G. Rau, E. K.-H. Lee, and H.-Y. Kee, Phys. Rev. Lett. \textbf{112}, 077204 (2014).

\bibitem{Singh2010} Y. Singh and P. Gegenwart, Phys. Rev. B \textbf{82}, 064412 (2010).
\bibitem{Liu2011} X. Liu, T. Berlijn, W.-G. Yin, W. Ku, A. Tsvelik, Y.-J. Kim, H. Gretarsson, Y. Singh, P. Gegenwart and J. P. Hill, Phys. Rev. B \textbf{83}, 220403 (2011).
\bibitem{Ye2012} F. Ye, S. Chi, H. Cao, B. C. Chakoumakos, J. A. Fernandez-Baca, R. Custelcean, T. F. Qi, O. B. Korneta, and G. Cao, Phys. Rev. B \textbf{85}, 180403(R) (2012).
\bibitem{Choi2012} S. K. Choi, R. Coldea, A. N. Kolmogorov, T. Lancaster, I. I. Mazin, S. J. Blundell, P. G. Radaelli, Y. Singh, P. Gegenwart, K. R. Choi, S.-W. Cheong, P. J. Baker, C. Stock, and J. Taylor, Phys. Rev. Lett. \textbf{108}, 127204 (2012).

\bibitem{Mazin2012} I. I. Mazin, H. O. Jeschke, K. Foyevtsova, R. Valent\'{i}, and D. I. Khomskii, Phys. Rev. Lett. \textbf{109}, 197201 (2012).
\bibitem{Foyevtsova2013} K. Foyevtsova, H. O. Jeschke, I. I. Mazin, D. I. Khomskii, and R. Valent\'{i}, Phys. Rev. B. \textbf{88}, 035107 (2013).

\bibitem{Johnson2015} R. D. Johnson, S. C. Williams, A. A. Haghighirad, J. Singleton, V. Zapf, P. Manuel, I. I. Mazin, Y. Li, H. O. Jeschke, R. Valent\'{i}, and R. Coldea, Phys. Rev. B \textbf{92}, 235119 (2015).
\bibitem{Cao2016} H. B. Cao, A. Banerjee, J.-Q. Yan, C. A. Bridges, M. D. Lumsden, D. G. Mandrus, D. A. Tennant, B. C. Chakoumakos, and S. E. Nagler, Phys. Rev. B \textbf{93}, 134423 (2016).
\bibitem{Kubota2015} Y. Kubota, H. Tanaka, T. Ono, Y. Narumi, and K. Kindo, Phys. Rev. B \textbf{91}, 094422 (2015).
\bibitem{Sears2015} J. A. Sears, M. Songvilay, K. W. Plumb, J. P. Clancy, Y. Qiu, Y. Zhao, D. Parshall, and Y.-J. Kim, Phys. Rev. B \textbf{91}, 144420 (2015).
\bibitem{Majumder2015} M. Majumder, M. Schmidt, H. Rosner, A. A. Tsirlin, H. Yasuoka, and M. Baenitz, Phys. Rev. B \textbf{91}, 180401(R) (2015).
\bibitem{Sandilands2016} L. J. Sandilands, Y. Tian, A. A. Reijnders, H.-S. Kim, K. W. Plumb, Y.-J. Kim, H.-Y. Kee, and K. S. Burch, Phys. Rev. B \textbf{93}, 075114 (2016).
\bibitem{Banerjee2016} A. Banerjee, C. A. Bridges, J.-Q. Yan, A. A. Aczel, L. Li, M. B. Stone, G. E. Granroth, M. D. Lumsden, Y. Yiu, J. Knolle, D. L. Kovrizhin, S. Bhattacharjee, R. Moessner, D. A. Tennant, D. G. Mandrus, and S. E. Nagler, Nature Materials \textbf{15}, 733 (2016).
\bibitem{Plumb2014} K. W. Plumb, J. P. Clancy, L. J. Sandilands, V. Vijay Shanker, Y. F. Hu, K. S. Burch, H.-Y. Kee, and Y.-J. Kim, Phys. Rev. B \textbf{90}, 041112(R) (2014).
\bibitem{Kim2015} H.-S. Kim, V.Vijay Shanker, A. Catuneau, and H.-Y. Kee, Phys. Rev. B \textbf{91}, 241110(R) (2015).
\bibitem{Zhou2016} X. Zhou, H. Li, J. A. Waugh, S. Parham, H.-S. Kim, J. A. Sears, A. Gomes, H.-Y. Kee, Y.-J. Kim, and D. S. Dessau, Phys. Rev. B \textbf{94}, 161106(R) (2016).
\bibitem{Koitzsch2016} A. Koitzsch, C. Habenicht, E. M\"{u}ller, M. Knupfer, B. B\"{u}chner, H. C. Kandpal, J. van den Brink, D. Nowak, A. Isaeva, and Th. Doert, Phys. Rev. Lett. \textbf{117}, 126403 (2016). 
\bibitem{Sandilands2015} L. J. Sandilands, Y. Tian, K. W. Plumb, Y.-J. Kim, and K. S. Burch, Phys. Rev. Lett. \textbf{114}, 147201 (2015).
\bibitem{Sandilands2016_2} L. J. Sandilands, C. H. Sohn, H. J. Park, S. Y. Kim, K. W. Kim, J. A. Sears, Y.-J. Kim, and T. W. Noh, Phys. Rev. B \textbf{94}, 195156 (2016).  
\bibitem{Hirobe2017} D. Hirobe, M. Sato, Y. Shiomi, H. Tanaka, and E. Saitoh, Phys. Rev. B \textbf{95}, 241112(R) (2017).
\bibitem{Yoshitake2016} J. Yoshitake, J. Nasu, and Y. Motome, Phys. Rev. Lett. \textbf{117}, 157203 (2016).
\bibitem{Nasu2016} J. Nasu, J. Knolle, D. L. Kovrizhin, Y, Motome, and R. Moessner, Nat. Phys. \textbf{12}, 912 (2016).
\bibitem{Nasu2017} J. Nasu, J. Yoshitake, and Y. Motome, Phys. Rev. Lett. \textbf{119}, 127204 (2017).
\bibitem{Winter2017} S. M. Winter, K. Riedl, P. A. Maksimov, A. L. Chernyshev, A. Honecker, Nat. Commun. \textbf{8}, 1152 (2017). 
\bibitem{Winter2018} S. M. Winter, K. Riedl, D. Kaib, R. Coldea, and R. Valent\'{i}, Phys. Rev. Lett. \textbf{120}, 077203 (2018). 
\bibitem{Glamazda2017} A. Glamazda, P. Lemmens, S.-H. Do, Y. S. Kwon, and K.-Y. Choi, Phys. Rev. B \textbf{95}, 174429 (2017).
\bibitem{Aoyama2017} T. Aoyama, Y. Hasegawa, S. Kimura, T. Kimura, and K. Ohgushi, Phys. Rev. B \textbf{95}, 245104 (2017).
\bibitem{Hasegawa2017} Y. Hasegawa, T. Aoyama, K. Sasaki, Y. Ikemoto, T. Moriwaki, T. Shirakura, R. Saito, Y. Imai, and K. Ohgushi, J. Phys. Soc. Jpn. \textbf{86}, 123709 (2017).
\bibitem{Leahy2017} I. A. Leahy, C. A. Pocs, P. E. Siegfried, D. Graf, S.-H. Do, K.-Y. Choi, B. Normand, and M. Lee, Phys. Rev. Lett. \textbf{118}, 187203 (2017).
\bibitem{Sears2017} J. A. Sears, Y. Zhao, Z. Xu, J. W. Lynn, and Y.-J. Kim, Phys. Rev. B \textbf{95}, 180411(R) (2017).
\bibitem{Baek2017} S.-H. Baek, S.-H. Do, K.-Y. Choi, Y. S. Kwon, A. U. B. Wolter, S. Nishimoto, J. van den Brink, and B. B\;{u}chner, Phys. Rev. Lett. \textbf{119}, 037201 (2017). 
\bibitem{Wolter2017} A. U. B. Wolter, L. T. Corredor, L. Janssen, K. Nenkov, S. Sch\;{o}necker, S.-H. Do, K.-Y. Choi, R. Albrecht, J. Hunger, T. Doert, M. Vojta, and B. B\;{u}chner, Phys. Rev. B \textbf{96}, 041405(R) (2017).
\bibitem{ZWang2017} Z. Wang, S. Reschke, D. H\;{u}vonen, S.-H. Do, K.-Y. Choi, M. Gensch, U. Nagel, T. R\~{oo}m, and A. Loidl, Phys. Rev. Lett. \textbf{119}, 227202 (2017).
\bibitem{Little2017} A. Little, L. Wu, P. Lampen-Kelley, A. Banerjee, S. Patankar, D. Rees, C. A. Bridges, J.-Q. Yan, D. Mandrus, S. E. Nagler, and J. Orenstein, Phys. Rev. Lett. \textbf{119}, 227201 (2017).
\bibitem{Zheng2017} J. Zheng, K. Ran, T. Li, J. Wang, P. Wang, B. Liu, Z.-X. Liu, B. Normand, J. Wen, and W. Yu, Phys. Rev. Lett. \textbf{119}, 227208 (2017).
\bibitem{Ponomaryov2017} A. N. Ponomaryov, E. Schulze, J. Wosnitza, P. Lampen-Kelley, A. Banerjee, J.-Q. Yan, C. A. Bridges, D. G. Mandrus, S. E. Nagler, A. K. Kolezhuk, and S. A. Zvyagin, Phys. Rev. B \textbf{96}, 241107(R) (2017). 

\bibitem{Sizyuk2016} Y. Sizyuk, P. W\"{o}lfle, and N. B. Perkins, Phys. Rev. B \textbf{94}, 085109 (2016).
\bibitem{Kim2016} H.-S. Kim and H.-Y. Kee, Phys. Rev. B \textbf{93}, 155143 (2016).
\bibitem{Winter2016} S. M. Winter, Y. Li, H. O. Jeschke, and Roser Valent\'{i}, Phys. Rev. B \textbf{93}, 214431 (2016).
\bibitem{Chaloupka2016} J. Chaloupka and G. Khaliullin, Phys. Rev. B \textbf{94}, 064435 (2016).
\bibitem{Lang2016} F. Lang, P. J. Baker, A. A. Haghighirad, Y. Li, D. Prabhakaran, R. Valent\'{i}, and S. J. Blundell, Phys. Rev. B \textbf{94}, 020407(R) (2016).
\bibitem{KT} R. S. Hayano, Y. J. Uemura, J. Imazato, N. Nishida, T. Yamazaki, and R. Kubo, Phys. Rev. B \textbf{20}, 850 (1979).
\bibitem{Kanigel2002} A. Kanigel, A. Keren, Y. Eckstein, A. Knizhnik, J. S. Lord, and A. Amato, Phys. Rev. Lett. \textbf{88}, 137003 (2002).
\bibitem{Dally2014} R. Dally, T. Hogan, A. Amato, H. Luetkens, C. Baines, J. Rodriguez-Rivera, and M. J. Graf, and S. D. Wilson, Phys. Rev. Lett. \textbf{113}, 247601 (2014).
\bibitem{Tanaka} R. Takeda, N. Kurita, and H. Tanaka, unpublished data.
\bibitem{Yasuoka1978} H. Yasuoka, R. S. Hayano, N. Nishida, K. Nagamine, T. Yamazaki, and Y. Ishikawa, Solid State Commun. \textbf{26}, 745 (1978).
\bibitem{demag} J. A. Osborn, Phys. Rev. \textbf{67}, 351 (1945). 
\bibitem{VESTA} K. Momma and F. Izumi, J. Appl. Crystallogr., \textbf{44,} 1272 (2011).
\bibitem{VASP} G. Kresse and J. Hafner, Phys. Rev. B \textbf{47}, 558 (1993).; G. Kresse and J. Hafner, \textit{ibid.} \textbf{49}, 14251 (1994).; G. Kresse and J. Furthm\"{u}ller, Comput. Mat. Sci. \textbf{6}, 15 (1996).; G. Kresse and J. Furthm\"{u}ller, Phys. Rev. B \textbf{54}, 11169 (1996).
\bibitem{Bernardini2013} F. Bernardini, P. Bonf\'{a}, S. Massidda, and R. De Renzi, Phys. Rev. B \textbf{87}, 115148 (2013).
\bibitem{Moller2013} J. S. M\"{o}ller, D. Ceresoli, T. Lancaster, N. Marzari, and S. J. Blundell, Phys. Rev. B \textbf{87}, 121108(R) (2013). 
\bibitem{OpneMX} T. Ozaki, Phys. Rev. B \textbf{67}, 155108 (2003).; T. Ozaki and H. Kino, Phys. Rev. B \textbf{69}, 195113 (2004).; T. Ozaki and H. Kino, Phys. Rev. B \textbf{72}, 045121 (2005).; K. Lejaeghere \textit{et al.}, Science \textbf{351}, aad3000 (2016).
\bibitem{Yaouanc} A. Yaouanc and P. Dalmas de R\'{e}otier, \textit{Muon Spin Rotation, Relaxation, and Resonance: Applications to Condensed Matter} (Oxford University Press, Oxford, 2011).
\bibitem{Yadav2016} R. Yadav, N. A. Bogdanov, V. M. Katukuri, S. Nishimoto, J. van den Brink, and L. Hozoi, Sci. Rep. \textbf{6}, 37925 (2016).
\bibitem{Banerjee2017} A. Banerjee, J. Yan, J. Knolle, C. A. Bridges, M. B. Stone, M. D. Lumsden, D. G. Mandrus, D. A. Tennant, R. Moessner, and S. E. Nagler, Science \textbf{356}, 1055 (2017).
\bibitem{Ran2017} K. Ran, J. Wang, W. Wang, Z.-Y. Dong, X. Ren, S. Bao, S. Li, Z. Ma, Y. Gan, Y. Zhang, J. T. Park, G. Deng, S. Danilkin, S.-L. Yu, J.-X. Li, and J. Wen, Phys. Rev. Lett. \textbf{118}, 107203 (2017).
\bibitem{Do2017} S.-H. Do, S.-Y. Park, J. Yoshitake, J. Nasu, Y. Motome, Y. S. Kwon, D. T. Adroja, D. J. Voneshen, K. Kim, T.-H. Jang, J.-H. Park, K.-Y. Choi, and S. Ji, Nat. Phys. \textbf{13}, 1079 (2017). 
\bibitem{Banerjee2017FI} A. Banerjee, P. Lampen-Kelly, J. Knolle, C. Balz, A. A. Aczel, B. Winn, Y. Liu, D. Pajerowski, J.-Q. Yan, C. A. Bridges, A. T. Savici, B. C. Chakoumakos, M. D. Lumsden, D. A. Tennant, R. Moessner, D. G. Mandrus, and S. E. Nagler, npj Quant. Mater. \textbf{3}, 8 (2018).
\bibitem{WWang2017} W. Wang, Z.-Y. Dong, S.-L. Yu, and J.-X. Li, Phys. Rev. B \textbf{96,} 115103 (2017).






\end{thebibliography}
\end{document}